\documentclass[twocolumn,showpacs,nofootinbib,pre]{revtex4}
\usepackage{amsmath}
\usepackage{latexsym}
\usepackage{epsfig}
\usepackage{graphicx}

\newcommand{\vep}{\varepsilon}
\newcommand{\be}{\begin{equation}}
\newcommand{\ee}{\end{equation}}
\newcommand{\bea}{\begin{eqnarray}}
\newcommand{\eea}{\end{eqnarray}}
\begin{document}

\title{Richardson diffusion in neurons}
\author{Alexander Iomin }
\affiliation{Department of Physics, Technion, Haifa, 32000, Israel}
\affiliation{\textit{Phys. Rev. E, Rapid Com. \textbf{100}, 010104(R)} (2019)}

\begin{abstract}
The dynamics of an initial wave packed affected by random noise is considered in the framework of a comb model.
The model is relevant to a diffusion problem in neurons
where the transport of ions can be accelerated by an external
random field due to synapse fluctuations. In the
present specific case, it acts as boundary conditions, which
lead to a reaction transport equation with multiplicative noise.
The temporal behavior of the mean squared
displacement  is estimated analytically,
and it is shown that the spreading of the initial wave packet
corresponds to Richardson diffusion.
\end{abstract}

\maketitle

\textit{Introduction.-}
Recent experimental investigations show
that transport of an initial wave packet can be accelerated inside
space-time disordered media \cite{segev1,LeKrFiSe12}.
In particular, hyper-diffusion (as a quantum realization of
Richardson diffusion \cite{richardson}) has been observed,
experimentally and numerically, and
explained theoretically \cite{KrLeFiSeWi12,Io15}.
It is reasonably to believe that this phenomenon
has a generic nature and takes place not only
in the wave dynamics, and results from
spatio-temporal characteristics of random fields.
Here, we show that behind Richardson diffusion in the comb models,
there is the same mechanism based on a
phenomenological statistical approach, discussed
in quantum mechanical observation of Richardson diffusion \cite{Io15}.
Dated back to work by Kolmogorov and Obukhov \cite{frisch,monin_yaglom},
it suggests this turbulent acceleration by means of a
Gaussian delta correlated noise \cite{obukhov}, added to the
dynamical system $\ddot{x}+V(t)=0$. In this case,
due to the noise term $V(t)$,  Richardson diffusion
\cite{richardson} takes place with the mean squared
displacement (MSD) $\langle{x}^2(t)\rangle\sim t^3$,
which is due to the diffusive spread of the velocity
profile $\langle\dot{x}^2(t)\rangle\sim t$.
We consider a diffusion problem in neurons in
the framework of a comb model and show that the transport
can be accelerated by an external random field, which, in the
present specific case, acts as a boundary condition.

\begin{figure}
\centering\includegraphics[width=8cm]{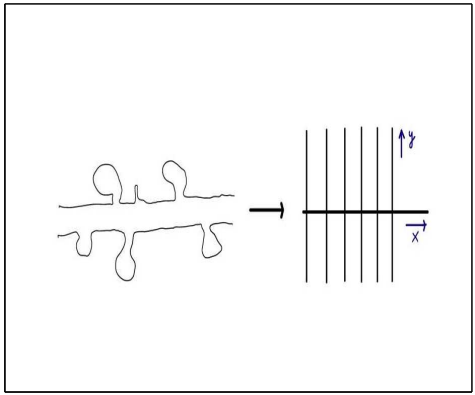}
\caption{(Color online) Mapping of a spine dendrite on a comb,
where fingers correspond to spines.
There is infinite number of $y$-channels continuously
distributed along the $x$ coordinates. In this case
at each $x$, the probability to enter to a finger
is $1/2$ (in either directions) and the probability to
move along the backbone is $1/2$ as well.
This relation between the  real
three dimensional Laplace operator and the Laplace operator
of the comb model \eqref{sfpe-comb}  was established
in Ref. \cite{IoZaPf16}.}
\label{fig:fig1}
\end{figure}

It has been shown in experimental and numerical studies that
the transport of inert particles along dendrite structure of neurons
corresponds to anomalous diffusion (namely subdiffusion),
when the temporal behavior of the MSD is of the power law $t^{\gamma}$,
where the transport exponent $0<\gamma<1$ depends on the
geometry and density of the dendritic spines \cite{santa1,santa2,ZeHo11}.
Dendritic spines are the basic functional units
in pre- and post synaptic activity of neurons
\cite{yuste1}, and further studies have shown that
the comb model can be used to describe the movement and binding dynamics
of particles, including  reaction transport
of ${\rm Ca}^{2+}$ ions inside the spines \cite{MI13,IM13,YuAbBa16}.
A comb model has been suggested as a simplified toy model, which
reflects this property of anomalous diffusion,
resulted from the geometry, which
%As it is seen in Fig.~\ref{fig:fig1}, the comb geometry
mimics geometry of spiny dendrites, such that the backbone
is the dendrite and the fingers are the spines,
see Fig.~\ref{fig:fig1}.

A special property of such geometry is reflected
in transport (diffusion) coefficients, such that
transport along the $x$ coordinate is possible along
the backbone  at $y=0$ only, while diffusion along the
$y$ coordinate is homogeneous. Therefore, the probability
to find a particle at the position
$(x,y)$ at time $t$ is determined by the probability distribution function
(PDF) $P=P(x,y,t)$, which is controlled by the Fokker-Planck
equation \cite{ArBa91}.
The corresponding equation in the dimensionless variables reads
\begin{equation}\label{sfpe-comb}
\partial_tP=\delta(y)\partial^2_xP+\partial_y^2P \, .
\end{equation}
For infinite combs, there is subdiffusion along the backbone
with the MSD of the order of $t^{\frac{1}{2}}$ \cite{WeHa86}.
This fractional diffusion in the comb
reflects a neuronal property of the power law adaptation,
which results in neuronal fractional differentiation,
observed experimentally \cite{LuHiSpFa08}, as well.
For a finite comb with finite length
$h$ of fingers, this subdiffusion takes place at
times $t<h$, and then it switches to normal diffusion at
$t>h$ \cite{FoBuCeVu13,IoZaPf16}.

It should be admitted that this multiscale dynamics
in the finite combs depends also on boundary
conditions at finite fingers-spines.
These boundary conditions are determined by unstable
synapses \cite{naama},  undergoing random
fluctuations\footnote{See this discussion in
Ref. \cite{naama} and references therein},
can be considered as a random nose at boundaries.
Eventually, we arrived at a simple model - comb model, whose
geometry mimics the neuron spiny dendrite and the boundary conditions
mimic the synapse random fluctuations.
These boundary conditions are defined as follows
\begin{equation}\label{sfpe-2}
\partial_yP(x,y,t)\left|_{y=h}\right.
-\partial_yP(x,y,t)\left|_{y=-h}\right. =W(x,t)\, .
\end{equation}
It is worth noting that the boundary conditions at $y=\pm h$
correspond to the same spine (or synapse). Therefore, $W(x,t)$
consists of two identical fluxes with the opposite directions.

\textit{Stochastic Fokker-Planck equation.-}
Therefore, the influence of the boundary
fluctuations on the particle transport (including reactions)
in neurons is studied in the framework
of the comb model \eqref{sfpe-comb} with
random boundary conditions for the fingers described in
Eq. \eqref{sfpe-2}.
Here we consider a multiplicative noise
$w(x,t)$ in the form $W(x,t)=w(x,t)\rho(x,t)$, where
$\rho(x,t)$ is a marginal
PDF, which determines transport along the backbone.
The distribution of $w(x,t)$ and its spatio-temporal characteristics
will be defined in the text in such a way that it will be
suitable for the MSD calculations.

The backbone transport is described by either marginal PDF
\begin{equation}\label{sfpe-3}
\rho(x,t)=\int_{-h}^{h}P(x,y,t)dy\, ,
\end{equation}
or by the backbone PDF $P(x,y=0,t)$.
Here we consider a diffusion process on the times $t>h$. In this case,
there is a simple relation between these PDFs $P(x,y=0,t)\sim
h\rho(x,t)$,
which reduces to the equality at the asymptotically large time scale,
$t\gg h$. As it follows from Ref. \cite{YuAbBa16},
this relation should be also true for random finger's
length with the finite mean length
of the order of $h$.

Performing integration with respect to $y$, one arrives at the stochastic
Fokker-Planck equation (SFPE)
\begin{equation}\label{sfpe-4}
\partial_t\rho(x,t)=h\partial_x^2\rho(x,t)+w(x,t)\rho(x,t)
\end{equation}
with the initial condition $\rho_0(x)=\rho(x,t=0)=\delta(x)$.
In the case of additive noise, this equation is also known as an
Edwards-Wilkinson equation \cite{EdWi82}.

Important caution here is that to avoid an avalanche, exponential
increasing of the number of transporting particles due to the random
reaction term $w(x,t)\rho(x,t)$, we impose the restriction condition,
which controls the total number of particles. In particular,
we can consider a conservation rule of the total number of particles
at every realization of the random noise $w(x,t)$.
It reads\footnote{Later, we shall suggest more realistic
Fisher-Kolmogorov-Petrovskii-Piskunov mechanism of
the reaction control. At the point we do not
specify a mechanism of such restriction.
However, if for every realization of
$w(x,t)$ and for any random walk of a particle,
the probability to find it inside the boundaries is one
then condition \eqref{sfpe-5} is fulfilled.
It is worth noting that the implementation of this
condition supposes also
free boundary conditions
at the dendrite-axon connection $x=x_{\rm d-a}\equiv X$, which
leads to a free nonzero current $j(X,t)$ from the dendrite
to the axon.
Therefore, integration of the equation with respect to $x$ yields
$\int_xw(x,t)\rho(x,t)dx=j(X,t)$. In this case,
there is no any restrictions of the random noise $w(x,t)$.
}
\begin{equation}\label{sfpe-5}
\int_x\rho(x,t)dx=1\, .
\end{equation}

The solution of the SFPE \eqref{sfpe-4} can be presented in the form of
the time ordered exponentials as follows
\begin{equation}\label{sfpe-6}
\rho(x,t)=\hat{T}\exp
\left[\int_0^t\left(h\partial_x^2+w(x,\tau)\right)d\tau\right]
\rho_0(x)\, ,
\end{equation}
where $\hat{T}$ is the time ordering operator, and
under this sign all values are commute.
Applying the Hubbard-Stratonovich transformation
for the second derivative, we present it as a shift operator
\begin{equation}\label{sfpe-7}
e^{ht\partial_x^2}=\int\prod_{\tau}\frac{d\lambda(\tau)}{\sqrt{4\pi/hd\tau}}
e^{-\frac{h}{4}\int_0^t\lambda^2(\tau)d\tau}
e^{h\partial_x\int_0^t\lambda(\tau)d\tau}\, .
\end{equation}
This yields solution \eqref{sfpe-6} in the form of the Feynman-Kac
path integral
\begin{multline}\label{sfpe-8}
\rho(x,t)=\int\prod_{\tau}\frac{d\lambda(\tau)}{\sqrt{4\pi/hd\tau}} \\
\times e^{-\frac{h}{4}\int_0^t\lambda^2(\tau)d\tau}
e^{\int_0^tw\left(x_{\tau}(t),\tau\right)d\tau}\delta\left(x(t)\right)\, ,
\end{multline}
where
\begin{equation}\label{sfpe-9}
x_{\tau}(t)=x+h\int_{\tau}^t\lambda(\tau)d\tau\, , \quad
x(t)=x+h\int_0^t\lambda(\tau)d\tau\, .
\end{equation}

We substitute this solution in the restriction
condition \eqref{sfpe-5}
and take into account the delta function for the
integration with respect to $x$.
The path integral is estimated by the extremum principal
Hamiltonian function, or action $S_{\rm e}$,
which yields Eq. \eqref{sfpe-5} as follows
\begin{equation}\label{sfpe-10}
F(t)e^{-S_{\rm e}(T)}=1\, .
\end{equation}
Here, the pre-factor $F(T)$ stands for the normalization condition
and compensates the exponential proliferation of particles\footnote{In fact
it is unknown complicated, random function.
However, due to this restriction condition,
the explicit form for this pre-factor is not important.}.
The extremum action is determined from the condition $\delta S(T)=0$,
where
\[S(T)=\int_0^TL(\dot{X},X,t)dt=
\int_0^T\left[\frac{1}{4h}\dot{X}^2-w(-X,t)\right]dt\, .\]
Here the velocity and the coordinate are
$\dot{X}=h\lambda(t)$ and $X=x+h\int_0^t\lambda(\tau)d\tau$,
correspondingly. The extremum action $S_{\rm e}(T)$
is determined by the Euler-Lagrange equation,
which corresponds to the velocity functional
\begin{equation}\label{sfpe-11}
\lambda(t)=2h\int_0^t\frac{\partial w(-X,\tau)}{\partial X}d\tau\, .
\end{equation}

\textit{Richardson diffusion.-}
Now the MSD $\langle\langle x^2(t)\rangle\rangle_{w}$,
averaged over all possible realizations of the random
force $f(x,t)=\frac{\partial w(-X,\tau)}{\partial( -X)}$ can
be estimated. Taking into account Eq. \eqref{sfpe-10}, we have
\begin{multline}\label{sfpe-12}
\langle\langle x^2(t)\rangle\rangle_{w}=
\Big\langle \int_{x}x^2\rho(x,t)dx {}\Big \rangle_{w} \\
=4h^2\int_0^Tdt\int_0^td\tau\int_0^Tdt'\int_0^{t'}d\tau'
\langle f(x,\tau)f(x,\tau')\rangle_w\, .
\end{multline}

Now we can suppose the correlation properties of the random noise,
in such a way that both $w(x,t)$ and $f(x,t)$ are Gaussian,
translational invariant in time and space, and delta correlated
in time, and their correlation functions $C_R(x,t;x't')=
\langle R(x,t)R(x',t')\rangle_w$ with $R=\binom{w}{f}$ are determined
by a spectral density $S(k)$ as follows
\begin{subequations}\label{sfpe-13}
\begin{align}
C_w(x,t;x',t')= & C_w(x-x')\delta(t-t') \nonumber \\
= & \int S(k)\cos[k(x-x')]dk\delta(t-t') \, , \label{sfpe-13a}  \\
C_f(x,t;x',t')= & C_f(x-x')\delta(t-t') \nonumber \\
= & \int k^2S(k)\cos[k(x-x')]dk\delta(t-t')\, . \label{sfpe-13b}
\end{align}
\end{subequations}
Taking into account correlation \eqref{sfpe-13b} in
Eq. \eqref{sfpe-12}, we arrive at Richardson diffusion \cite{obukhov}
with the MSD
\begin{equation}\label{sfpe-14}
\langle\langle x^2(t)\rangle\rangle_{w}=2h^2Dt^3\, ,
\end{equation}
where $D=\int k^2S(k)dk$ is a transport coefficient

%\textit{Backbone's boundary conditions}

\textit{Reaction front propagation.-}
Important part of the analysis is the restriction, or
control of the number particles.
It is a common
statement for any mechanism of the control of the number of diffusive
particles, which prevents to the uncontrolled exponential increasing
of the particle's number due to the random reaction term $w(x,t)\rho(x,t)$
in SFPE \eqref{sfpe-4}. However, the number conservation condition \eqref{sfpe-5} is too strong, as admitted above,
and in general case, the number of particles
cannot be conserved due to reactions. In this case, a more
reasonable and realistic mechanism of the reaction control is due to a
FKPP (Fisher-Kolmogorov-Petrovskii-Piskunov) term, which should be inserted
in SFPE \eqref{sfpe-4}. The latter now reads
\begin{equation}\label{fkpp-1}
\partial_t\rho=h\partial^2_x\rho+w(1-\rho)\rho\, .
\end{equation}
This stochastic reaction-transport equation can be important
for understanding of translocation waves of ${\rm Ca^{2+}}$
ions in spiny dendrites, studied in the framework of the FKPP
scheme \cite{ErBr10,IM13}.
Stochastic FKKP term in Eq. \eqref{fkpp-1},
is a random generalization of a standard FKPP reaction term
$\rho(1-\rho)$, which is widely used in reaction transport equations
\cite{MeFeHo10}.

In this nonlinear case, the exact analytical treatment is not possible
anymore, and we apply an analytical approximation to estimate the overall
velocity of the reaction front propagation without resolving
the exact shape of the front. The method is based on a
hyperbolic scaling of space-time variables
$(x,t)$ by a small parameter $\vep$.
Following Ref. \cite{freidlin}, we introduce this parameter
$\vep$ at the derivatives.
To this end, we res-scale
$x\rightarrow x/\vep$ and $t\rightarrow t/\vep$,
and for the marginal PDF we have
$\rho(x,t)\rightarrow \rho^{\vep}(x,t)=
\rho\left(\frac{x}{\vep},\frac{t}{\vep}\right)$.
We look for the asymptotic solution in the form of the Green approximation
\begin{equation}\label{fkpp-2}
\rho^{\vep}(x,t)=exp\left[-S^{\vep}(x,t)/\vep\right]\, .
\end{equation}
The main strategy of the implication of this construction is the
limit $\vep\rightarrow 0$ that yields the asymptotic solution
at finite values of $x$ and $t$, such that $\rho^{\vep}(x,t)$
is not vanishing only when $S^{\vep}(x,t)=0$.
Therefore, expression \eqref{fkpp-2} is an extremum solution,
which determines the position of the reaction spreading front.
Substituting solution \eqref{fkpp-2} in Eq. \eqref{fkpp-1},
scaled by $\vep$, and taking limit $\vep\rightarrow 0$, we obtain
that $S^{\vep}(x,t)$ is an extremum solution:
$\lim_{\vep\to 0}S^{\vep}(x,t)=S_{\rm e}(x,t)$, which is
the extremum action, or the Hamilton's principal function.
It is determined by the Hamilton-Jacobi equation
\begin{equation}\label{fkpp-3}
-\partial_tS_{\rm e}=h\left(\partial_xS_{\rm e}\right)^2+w(x,t)\, .
\end{equation}
Taking into account that $-\partial_tS_{\rm e}=H$ is Hamiltonian and
$\partial_xS_{\rm e}=p$ is the momentum, we arrive at
the particle dynamics in a random noise potential with the Hamiltonian
$H=hp^2+w(x,t)$.

Further analysis differs from the standard approach of Ref.~\cite{freidlin},
where a particle is free, but here it is in a random potential.
Eventually, we arrived at the same mechanism of turbulent diffusion,
considered above in Eq. \eqref{sfpe-11}. Therefore, in the framework
of the Hamiltonian approach, the overall velocity of
the reaction front reads
\begin{equation}\label{fkpp-4}
V=\dot{x}=2hp=2h\int_0^t f(x,\tau)d\tau\, .
\end{equation}
The correlation properties of the random force is described by
Eq. \eqref{sfpe-13b}, and
we obtain the mean squared velocity (MSV) as follows
$\langle V^2(t)\rangle_{w}=4h^2Dt$,
which corresponds to Richardson diffusion with the MSD
$\langle x^2(t)\rangle_{w}=2h^2Dt^3$. It coincides  exactly with
the MSD in Eq. \eqref{sfpe-14}. Note that in both cases, $h$ is accounted as
the particle inverse mass.

\textit{Discussion.-}
We obtained that the SFPE \eqref{sfpe-4} with
the restriction condition \eqref{sfpe-5}, or the FKPP mechanism
controlling the number of transporting particles,
describes a reaction transport
process  in the presence of random boundary conditions. The latter
plays a role of accelerator mechanism of reaction transport and
leads to Richardson diffusion. Important condition
of the applicability of the SFPE \eqref{sfpe-4} for the transport
inside the comb model considered as a toy model of spiny dendrites,
is the long time asymptotics. In this case the transport
corresponds to  normal diffusion. Eventually, it corresponds to
a kind of Edwards-Wilkinson equation, where the random term is
a multiplicative noise. However, this equation does not describe
the initial time dynamics, which is important as well for the
time scale $t<h$. In this case the underlying kinetics inside
the backbone-dendrite is subdiffusion, due to the relation
in the Laplace space $\tilde{P}(x,y=0,s)=\sqrt{s}\tilde{\rho}(x,s)$.
This case leads to essential difficulties of the analysis and
can be important issue for future studies.

In conclusion, it should be admitted that an important motivation
of the research is possible experimental studies of transport
inside neurons, including artificial neurons \cite{art-n1}.
Another interesting possibility relates to experimental
investigations of reaction transport in
a micro-fluidic device of the comb geometry \cite{DePf12-15,IoZaPf16}
with the boundary control of fingers.

This research was supported by the Israel Science Foundation
(ISF-931/16).

\end{document}